\documentclass[conference,10pt]{IEEEtran}
\usepackage{amsmath}
\usepackage{amssymb}
\usepackage{amsfonts}
\usepackage{graphicx}
\usepackage{enumerate}
\usepackage{bbm}
\usepackage{subfig}
\usepackage{mathtools} % Added by Zheda
\usepackage{cite}
\usepackage{url}
\usepackage{array}
\usepackage{verbatim}
\usepackage{comment}
\usepackage{stfloats}
\usepackage{hyperref}

\usepackage{color}
\definecolor{red}{rgb}{1,0,0}
\definecolor{blue}{rgb}{0,0,1}

\usepackage[font=small,skip=-5pt]{caption}
\IEEEoverridecommandlockouts
\begin{document}
\title{Non-Orthogonal Multiple Access for mmWave Drones with Multi-Antenna Transmission}
\author{
\IEEEauthorblockA{Nadisanka~Rupasinghe\IEEEauthorrefmark{1}, Yavuz~Yap{\i}c{\i}\IEEEauthorrefmark{1}, \. {I}smail~G\"uven\c{c}\IEEEauthorrefmark{1}, and Yuichi Kakishima\IEEEauthorrefmark{2}, 
\IEEEauthorblockA{\IEEEauthorrefmark{1}Department of Electrical and Computer Engineering, North Carolina State University, Raleigh, NC}
\IEEEauthorblockA{\IEEEauthorrefmark{2}DOCOMO Innovations Inc., Palo Alto, CA}
{\tt \{rprupasi, yyapici, iguvenc\}@ncsu.edu,  kakishima@docomoinnovations.com}}%
\thanks{This research was supported in part by the U.S. National Science Foundation under the grant CNS-1618692.}
}
\maketitle

\begin{abstract}

Unmanned aerial vehicles (UAVs) can be deployed as aerial base stations (BSs) for rapid establishment of communication networks during temporary events and after disasters. Since UAV-BSs are low power nodes, achieving high spectral and energy efficiency are of paramount importance. In this paper, we introduce non-orthogonal multiple access (NOMA) transmission for millimeter-wave (mmWave) drones serving as flying BSs at a large stadium potentially with several hundreds or thousands of mobile users. In particular, we make use of multi-antenna techniques specifically taking into consideration the physical constraints of the antenna array, to generate directional beams. Multiple users are then served within the same beam employing NOMA transmission. If the UAV beam can not cover entire region where users are distributed, we introduce beam scanning to maximize outage sum rates. The simulation results reveal that, with NOMA transmission the spectral efficiency of the UAV based communication can be greatly  enhanced compared to orthogonal multiple access (OMA) transmission. Further, the analysis shows that there is an optimum transmit power value for NOMA beyond which outage sum rates do not improve further.

\end{abstract}


\section{Introduction} 

% In \cite{Ding17PoorRandBeamforming}, a random beamforming approach for mmWave NOMA networks are proposed where a single BS is supporting users who are randomly distributed across a disk with respect to a homogeneous Poisson point process (HPPP). In that work, the BS generates a random beam which creates a wedge-shaped sector, and users within this wedge-shaped sector are then considered for NOMA transmission. In particular, two users with different effective channel gains are selected for NOMA transmission within each beam.
Non-orthogonal multiple access (NOMA) is a promising technology for the next generation communication systems~\cite{Saito13VTC,3GPP16NOMA_LTE} due to its high spectral efficiency achieved through combining superposition coding at the transmitter with successive interference cancellation (SIC) at the receivers. In contrast to the conventional orthogonal multiple access (OMA) schemes (e.g., time-division multiple access (TDMA)), NOMA simultaneously serves multiple users in non-orthogonal resources, %the same degrees of freedom 
by separating the users in the power domain. 

%\textcolor{red}{degree of freedoms feels confusing since it may mean different things; check if update looks OK}% fundamentally different from conventional orthogonal multiple access (MA) schemes, as in NOMA multiple users are encouraged to transmit at the same time, code and frequency, but with different power levels.

In \cite{Saito13VTC}, a system-level performance evaluation based on 3GPP settings is carried out to identify achievable performance gains with NOMA over orthogonal frequency division multiple access (OFDMA). Achievable performance with NOMA when users are randomly deployed is investigated in \cite{Ding14NOMAfor5G_RandUsers} considering two different criteria: 1) when each user has a targeted rate based on their quality-of-service (QoS) requirements; and 2) opportunistic user rates based on their channel conditions. Provided that system parameters are appropriately chosen, better rate performance can be observed with NOMA compared to its orthogonal counterpart under both criteria. In \cite{Timotheou15PAforFairness}, a power allocation strategy for NOMA transmission is discussed considering user fairness in the DL data transmission. A cooperative NOMA strategy is proposed in \cite{Ding15Coop_NOMA} where strong users act as relays for weaker users with poor channel conditions to enhance their reception reliability.

% \textcolor{red}{check this sentence; you first refer to first criteria, but then refer to both; do we not need to choose the system parameters appropriately for second one??}

In \cite{Ding16MIMO_NOMA}, multiple-input-multiple-output (MIMO) techniques are introduced to NOMA transmission along with user pairing and power allocation strategies to enhance MIMO-NOMA performance over MIMO-OMA. A general MIMO-NOMA framework applicable to both downlink (DL) and uplink (UL) transmission is proposed in \cite{Ding16Schober_MIMO_NOMA} by considering signal alignment concepts. In \cite{Ding17PoorRandBeamforming}, a random beamforming approach for millimeter (mmWave) NOMA networks is proposed. To achieve power domain user separation, effective channel gains of users which depend on the angle offset between the randomly generated base station (BS) beam and user locations are considered. Two users are then served simultaneously within a single BS beam by employing NOMA techniques.

% When m is small, the mth user’s channel condition is poor, and the data rate supported by this user’s channel is also small. Therefore, the spectral efficiency of conventional MA is low since the bandwidth allocated to this user cannot be accessed by other users.

In this paper, we consider a scenario where an unmanned aerial vehicle BS (UAV-BS) is deployed to provide coverage over a large stadium/concert with thousands of mobile users distributed on the ground. In order to enhance the spectral efficiency, we introduce NOMA transmission at UAV-BS where the mobile traffic can be offloaded. In particular, using a multi-antenna array consisting of single radio-frequency (RF) chain, UAV-BS generates a directional beam. There can be several mobile stations (MS) that may fall within one such beam with different angle offsets with respect to the boresight direction of the beam. Then, the power domain user separation is achieved by considering effective channel gains of those users which depend on the angle offsets from the boresight of the beam, distance to the UAV-BS and small scale fading. We define targeted data rates for users based on their quality-of-service (QoS) \cite{Ding14NOMAfor5G_RandUsers} requirements and evaluate NOMA performance by analyzing outage sum rates. 

We focus on identifying optimum placement for a UAV-BS by optimizing its altitude based on outage sum rates. Due to the physical constraints of the antenna array there can be some altitudes at which all MSs on the ground can not be covered from the UAV-BS beam. In such a situation beam scanning is proposed to maximize outage sum rates. Further, our analysis sheds light on how to improve energy efficiency of UAV-BS while achieving targeted rates for users. We realize this by identifying an optimum transmit power value for NOMA beyond which outage sum rates do not improve further.  

The rest of the paper is organized as follows. Section~\ref{sec:Sys_Model} captures the system model considered for DL NOMA transmission with single UAV-BS. Formulation of NOMA transmission strategy in the DL considering a single UAV-BS beam is discussed in detail in Section~\ref{Sec:NOMA_Transmission}. In Section~\ref{sec:Numerical_results} numerical results for UAV NOMA/OMA DL transmission are presented. Finally, Section~\ref{sec:conclusion} provides concluding remarks.

\section{System Model} \label{sec:Sys_Model}

\begin{figure}[!t]
\begin{center}
\includegraphics[width=0.45\textwidth]{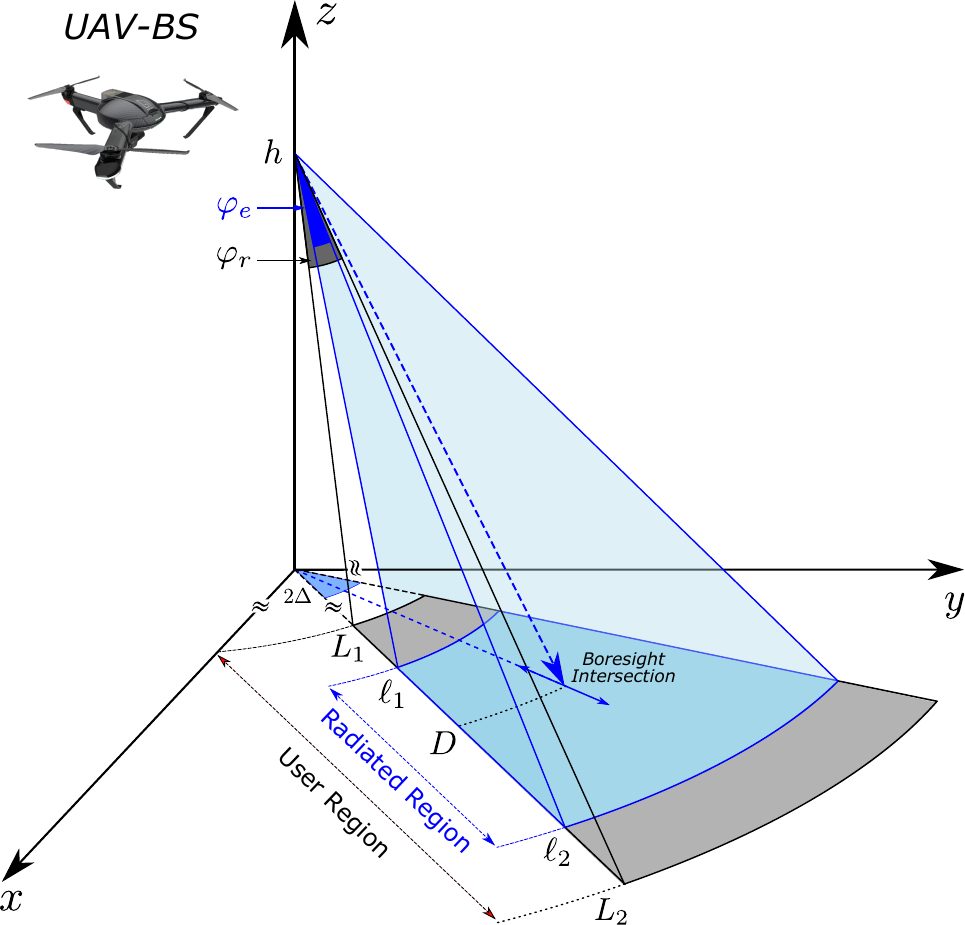}
\end{center}
\caption{3D footprint of the beam generated by UAV-BS partially covering the \emph{user region}, representing e.g., a sector of a stadium where mobile users are located. Here $\varphi_e$ captures the vertical beamwidth.}
\label{fig:footprint}
\vspace{0.25 em}
\end{figure}

We consider a mmWave-NOMA transmission scenario where a UAV-BS equipped with an $M$~element antenna array is serving mobile users in the DL and each user has a single antenna. We assume that all the users to be served by NOMA transmission lie inside a specific \emph{user region} as shown in Fig.~\ref{fig:footprint}, and are represented by the index set $\mathcal{N}_{\rm U} = \{1,2,\ldots K\}$. We also assume that the user region of interest may or may not be fully covered by a 3-dimensional (3D) beam generated by the UAV-BS depending on the specific geometry of the environment and 3D antenna radiation pattern. In Fig.~\ref{fig:footprint}, a particular scenario is depicted, where only part of the user region, which is labeled as \textit{radiated region}, is being covered by the UAV-BS beam. The user region is identified by inner-radius $L_1$, outer-radius $L_2$, and $2\Delta$ which is the fixed angle within the projection of azimuth beamwidth of the antenna pattern onto the $xy$-plane, as shown in Fig.~\ref{fig:footprint}. Similarly, the radiated region is described by the inner-radius $\ell_1$, outer-radius $\ell_2$, and the angle $2\Delta$. Note that it is possible to generate various stadium (and potentially other, e.g., concert, traffic jam, etc.) scenarios by modifying these parameters. For example, larger $L_1$ may correspond to a \textit{sports event} where users are only allowed to use the available seats on the tribunes while smaller $L_1$ may represent an \textit{event} where users can also be present on the ground as well as the tribunes.

\subsection{User Distribution and mmWave Channel Model} 
We assume that mobile users are randomly distributed following a homogeneous Poisson point process (HPPP) with density $\lambda$ \cite{Haenggi05Stochastic_Geo}. Hence, the number of users is Poisson distributed, i.e., $\textrm{P}(K \textrm{ users in the user region})\,{=}\, \frac{\mu^K e^{{-}\mu}}{K!} $ with $\mu\,{=}\,(L_2^2\,{-}\,L_1^2)\Delta \lambda$. The channel $\textbf{h}_k$ between the $k$-th user in the user region and the UAV-BS is given as
\begin{align} \label{k_UE_original_channel}
\textbf{h}_k = \sqrt{M} \sum \limits _{p=1}^{N_{\rm P}} \frac{\alpha_{k,p} \textbf{a}(\theta_{k,p})}{\sqrt{\textrm{PL}_k}},
\end{align} where $N_{\rm P}$, $\alpha_{k,p}$ and $\theta_{k,p}$ represent the number of multi-paths, gain of the $p$-th path which is complex Gaussian distributed with $\mathcal{CN}(0,1)$, and angle-of-departure (AoD) of the $p$-th path, respectively, and $\textbf{a}(\theta_{k,p})$ is the steering vector corresponding to AoD $\theta_{k,p}$. The path loss (PL) between $k$-th mobile user and UAV-BS is captured by $\textrm{PL}_k$. Without any loss of generality, we assume that all the users have line-of-sight (LoS) paths since UAV-BS is hovering at relatively high altitudes and the probability of having scatterers around UAV-BS is very small. Furthermore, as discussed in \cite{Ding17PoorRandBeamforming, Lee16mmWaveChannel}, the effect of LoS link over mmWave frequency bands is dominant as compared to those of the NLoS links. Hence, it is reasonable to assume a single LoS path for the channel under consideration, and \eqref{k_UE_original_channel} accordingly becomes:
\begin{align} \label{k_UE_modified_LoS_channel}
\textbf{h}_k = \sqrt{M} \frac{\alpha_k \textbf{a}(\theta_k)}{\sqrt{\textrm{PL}_k}}.
\end{align}

\subsection{Coverage of User Region} \label{Sec:Behavior_Vert_Angle}
\begin{figure}[!t]
\begin{center}
\includegraphics[width=0.45\textwidth]{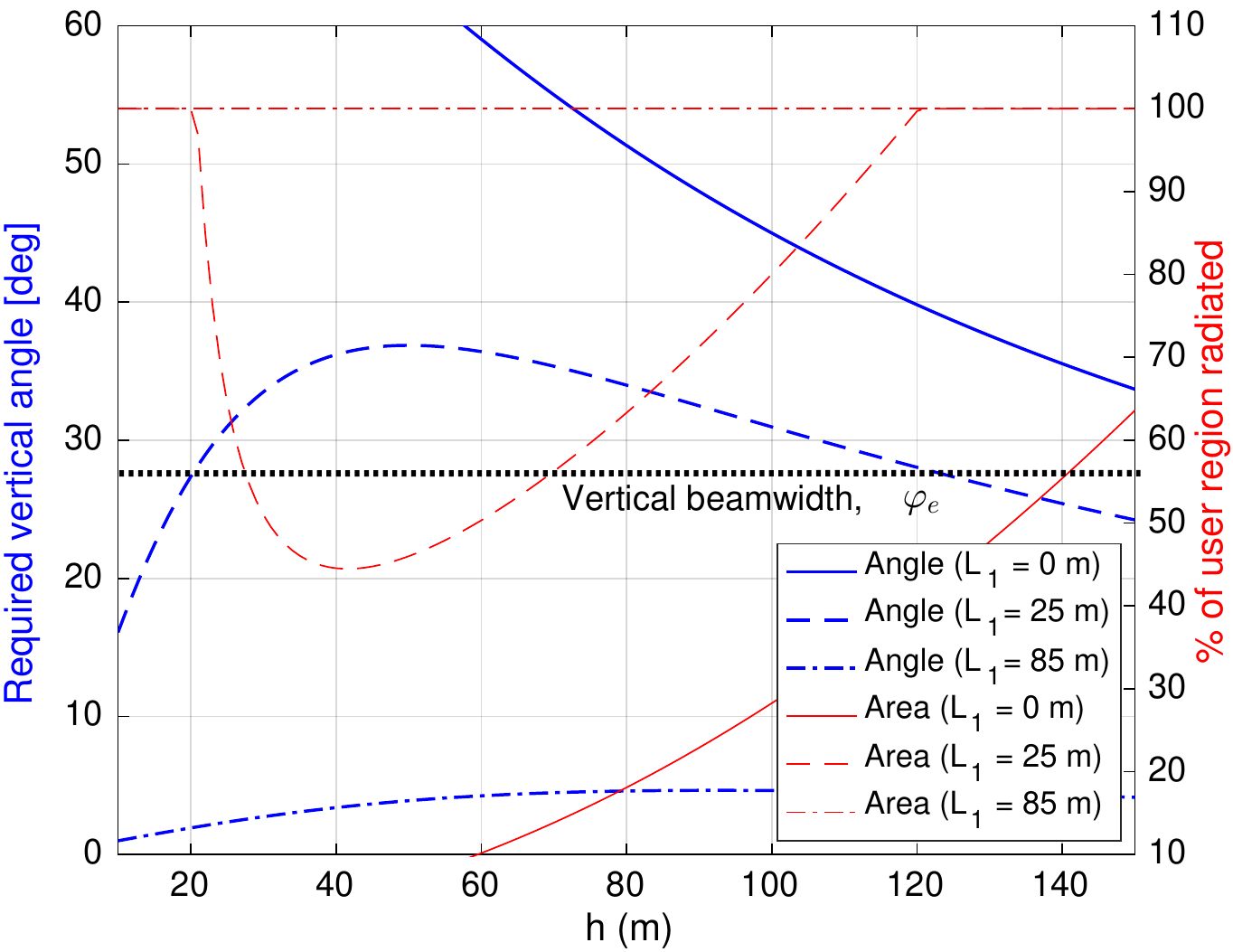}
\end{center}
\caption{Required vertical angle $\varphi_r$ to cover user region and percentage of user region radiated with vertical beamwidth $\varphi_e = 28^{\circ}$ at different altitudes for different geometries ($L_2 = 100$~m).}
\label{fig:VertAngle}
\end{figure}

Our primary goal in this paper is to find an optimal altitude value for the UAV operation such that the outage sum rate of the users in the predetermined user region is being maximized under NOMA transmission strategy. While doing that, it may not be possible to cover the entire user region since the vertical beamwidth $\varphi_e$ of the antenna radiation pattern of UAV-BS cannot be made arbitrarily large. To gain more insight into this, we depict the vertical angle $\varphi_r$, which is required to cover the entire user region, in Fig.~\ref{fig:VertAngle} for varying operation altitudes and user location scenarios. We also show in Fig.~\ref{fig:VertAngle}, the percentage of area radiated within the user region with the available vertical beamwidth $\varphi_e$. 

When the users are located everywhere in the stadium with $L_1\,{=}\,0$~m, the required vertical angle $\varphi_r$ is observed to decrease monotonically with increasing altitude, as shown in Fig.~\ref{fig:VertAngle}, as intuitively expected. However, when the inner-radius for user region is taken to be $L_1\,{=}\,25$~m, which may be the case when there is a big stage in the middle of the stadium during a concert setting, the required vertical angle $\varphi_r$ is increasing for relatively lower altitudes, and then starts to decrease. Because of this non-monotonic convex type behavior, practical vertical beamwidth $\varphi_e$ values of the transmitting UAV-BS antenna array may not be sufficiently large to cover the entire user region. For this particular case of $L_1\,{=}\,25$~m, the radiated region for a given vertical beamwidth of $\varphi_e\,{=}\, 28^{\circ}$ is shown to fall short of the entire user region over the altitude range of $h\,{\in}\,[21,120]$~m where $\varphi_e\,{<}\,\varphi_r$. Indeed, this altitude range is particularly important for the UAV-BS operation, since hovering at lower altitudes ($h\,{<}\,20$~m) is not recommended due to safety issues while higher altitudes ($h\,{>}\,120$~m) are typically restricted due to  regulations~\cite{FAARule}. 

\subsection{Beam Scanning over User Region}\label{Sec:Beam_Scanning}
When the physically radiated region is smaller than the desired user region, it may matter which portion of the entire user region should be covered. By moving the boresight intersection point radially forward and backward as shown in Fig.~\ref{fig:footprint}, it is possible to change the average path loss and radiated area size, both of which (among some other factors) can affect the sum rate significantly. It is therefore of particular interest to search for an optimal coverage within the user region for a given vertical beamwidth when $\varphi_e\,{<}\,\varphi_r$.      

We assume that the distance to the boresight intersection point of the beam from the origin is represented by $D$, as in Fig.~\ref{fig:footprint}. Keeping the radiated region fully inside the user region, we define $D_1$ and $D_2$ to be the two reasonable extreme values of $D$ where the inner-most and the outer-most portions of the user region are being covered, respectively. As a result, $D_1$ corresponds to the radiated region where $\ell_1\,{=}\,L_1$ and $\ell_2\,{<}\,L_2$, and $D_2$ corresponds to the scenario of $\ell_1\,{>}\,L_1$ and $\ell_2\,{=}\,L_2$. 

The beam scanning strategy therefore aims to find the optimum boresight intersection point $D^*$ such that the NOMA sum rates are maximized, and is formulated at a given altitude value of $h = \bar{h}$ as follows 
\begin{align} \label{eq:Opt_D}
D^{*} = \arg \max \limits _{(D_1 \leq D \leq D_2)} R^{\textrm{NOMA}},
\end{align} 
where $D_1$ and $D_2$ are $D_1\,{=}\,\bar{h} \tan \left(\tan^{{-}1}(L_1/\bar{h}) \,{+}\, \varphi_e/2 \right)$ and $D_2\,{=}\,\bar{h} \tan \left(\tan^{{-}1}(L_1/\bar{h}) \,{-}\, \varphi_e/2 \right)$ using the geometry of Fig.~\ref{fig:footprint}, and $R^{\textrm{NOMA}}$ is the NOMA sum rate. Note that, the optimum boresight intersection point $D^*$ can also be used to obtain the optimum tilting angle of UAV-BS in the DL transmission via the geometry of Fig.~\ref{fig:footprint}.

\section{NOMA Transmission with Single UAV Beam}
\label{Sec:NOMA_Transmission}

\begin{figure}[!t]
\begin{center}
\includegraphics[width=0.35\textwidth]{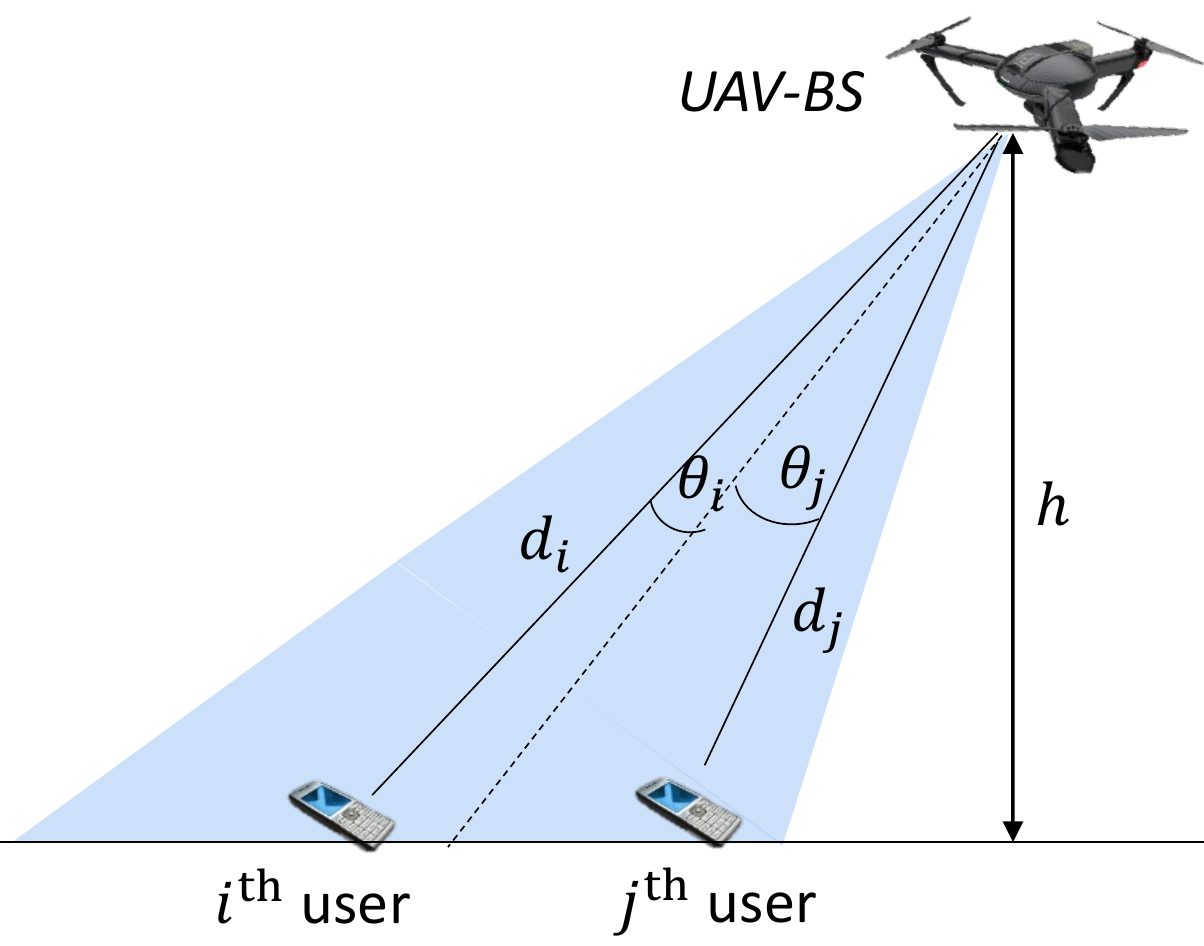}
\end{center}
\caption{Example scenario with two users are being served within single beam with NOMA. Effective channel gains of users depend on angle offset from the boresight (i.e., $\theta_k ,\, k \in \mathcal{N}_\textrm{U}$), distance to the UAV-BS (i.e., $d_k,\, k \in \mathcal{N}_\textrm{U})$, and path gain (i.e., $\alpha_k,\, k \in \mathcal{N}_\textrm{U})$.}
\label{fig:NOMA_Multi_UE_within_Beam}
\end{figure}

In this section, we study how to realize NOMA transmission for drone based communication scenario captured in Section~\ref{sec:Sys_Model}. In particular, as shown in Fig.~\ref{fig:NOMA_Multi_UE_within_Beam}, NOMA transmission is employed to serve multiple MSs within the same UAV beam (generated with single RF chain). We focus on outage sum rate performance in this study, since our goal here is to evaluate the possibility of achieving targeted rates for users participating in NOMA transmission.

To order users within the user region for NOMA transmission, we consider effective channel gains that are calculated by the MSs with respect to beam $\textbf{b} = \textbf{a}(\bar{\theta})$ transmitted by the UAV-BS. We assume UAV-BS generates beams sequentially one after the other such that the whole stadium can be covered. Hence, $\bar{\theta}$ the AoD of the beam  $\textbf{b}$ can take values from $0$ to $2 \pi$. Since the system is symmetric, analysis for a particular $\bar{\theta}$ is applicable to any $\bar{\theta} \in \left\lbrace 0, 2\pi \right\rbrace $. 

When the beam \textbf{b} is generated by the UAV-BS in a particular $\bar{\theta}$ direction, users fall within the user region (determined as discussed in Section~\ref{sec:Sys_Model} and with respect to the direction $\bar{\theta}$) are asked to measure their effective channel gains, $|\textbf{h}_k^{\rm H}\textbf{b}|^2, \ k \in \mathcal{N}_\textrm{U} $. The effective channel gain of the $k$-th user in the user region can be given as: \begin{align} \label{eq:Eff_channel_gain}
|\textbf{h}_k^{\rm H}\textbf{b}|^2 &= M \frac{|\alpha_k|^2 |\textbf{b}^{\rm H}\textbf{a}(\theta_k)|^2}{\textrm{PL}_k}
\nonumber \\
&= \frac{|\alpha_k|^2 M}{\textrm{PL}_k} \left| \frac{ \sin \left( \frac{\pi M (\sin \bar{\theta}-\sin \theta_k)}{2} \right)}{  M\sin \left( \frac{\pi (\sin \bar{\theta}-\sin \theta_k)}{2} \right)}\right|^2
\end{align} assuming critically spaced antenna array. These effective channel gains are fed back to UAV-BS. Next, we study how these effective channel gains are utilized for NOMA transmission.

\subsection{User Selection and DL Data Transmission for NOMA} \label{Sec:NOMA_User_Selection_Data_Transmission}

By considering effective channel gains of all users in the user region, UAV-BS first orders users in the ascending order with respect to their channel gains. To avoid use of tedious re-indexing, we assume without loss of generality that the users in set $\mathcal{N}_{\rm U}$ are already indexed in order of smaller to larger effective channel gains. Ordered effective channel gains of users can then be given as follows: \begin{align} \label{Sorted_Ch_gain}
|\textbf{h}_1^{\rm H}\textbf{b}|^2 \leq \cdots \leq |\textbf{h}_{K}^{\rm H}\textbf{b}|^2.
\end{align} Since NOMA exploits power domain in its operation, user with the worst channel condition is allocated the highest power whereas the user with the best channel condition is allocated the smallest power. Based on the user ordering in \eqref{Sorted_Ch_gain} and following the principal of NOMA, the power allocation coefficients of users can be ordered as, $\beta_1 \geq \cdots \geq 
\beta_K$ with the constraint $\sum \limits_{k = 1}^{K} \beta_k^2 = 1$.

With this power allocation, UAV-BS generates the DL transmitting signal $\textbf{x}$ by superposing messages of users participating in NOMA in the power domain as follows: \begin{align} \label{eq:Tx_signal}
\textbf{x} =\sqrt{P_{\rm Tx}}\textbf{b}\sum \limits_{k = 1}^{K} \beta_k s_{k}~,
\end{align} where $P_{\rm Tx}$ and $s_{k}$  are the total transmit power and the data for $k$-th user, respectively.
% As mentioned previously, since $|\textbf{h}_i^H\textbf{p}|^2 < |\textbf{h}_1^j\textbf{p}|^2$, for realizing NOMA transmission, $\beta_i \geq \beta_j$ with $\beta_i^2 + \beta_j^2 = 1$.
Based on the DL transmitted signal $\textbf{x}$ in \eqref{eq:Tx_signal}, user~$k \, \in \mathcal{N}_\textrm{U} $ will receive the following observation: \begin{align}\label{eq:k_th_user_Observation}
y_{k}= \textbf{h}_{k}^{\rm H} \textbf{x} +  N_0 = \sqrt{P_{\rm Tx}}\textbf{h}_{k}^{\rm H} \textbf{b}\sum \limits_{k = 1}^{K} \beta_k s_{k} + N_0,
\end{align} where $N_0$ is the additive noise. Now, considering successive interference cancellation (SIC), $k$-th user decodes and then subtracts the signals intended to first $(k-1)$ users having smaller effective channel gains in \eqref{Sorted_Ch_gain}. At the $k$-th user, data of $m$-th user, $1 \leq m \leq (k-1)$ will be detected with the following signal-to-interference-plus-noise ratio (SINR) \begin{align} \label{eq:SINR_mk_th_user}
\textrm{SINR}_{m({k})} = \frac{P_{\rm Tx}|\textbf{h}_{k}^{\rm H}\textbf{b}|^2 \beta_{m}^2}{P_{\rm Tx} \sum \limits_{l = (m+1)}^{K}|\textbf{h}_{k}^{\rm H}\textbf{b}|^2 \beta_{l}^2 + N_0}.
\end{align} As discussed previously, our focus in this work is to achieve guaranteed rates for users participating in NOMA. Hence, we consider $R_k^{\rm (T)}, \, k \in \mathcal{N}_\textrm{U}$ as the target rate for $k$-th user. Now, with the given target rate of $m$-th user $R_m^{\rm (T)}$, if $\textrm{SINR}_{m({k})} \geq \epsilon_m^{\rm (T)}$ in \eqref{eq:SINR_mk_th_user} where $\epsilon_m^{\rm (T)} = 2^{R_m^{\rm (T)}} - 1$, user $k$ can successfully remove $m$-th user data in \eqref{eq:k_th_user_Observation} and SIC can be carried out until his own message is decoded. Further, $k$-th user considers data for $(K-k)$ users having larger effective channel gains compared to his own channel gain as additive noise. The resulting SINR at $k$-th user can be given as: \begin{align} \label{eq:SINR_k_th_user}
{\rm SINR}_{k} = \frac{P_{\rm Tx}|\textbf{h}_{k}^{\rm H}\textbf{b}|^2 \beta_{k}^2}{P_{\rm Tx}  \sum \limits_{l = (k+1)}^{K}|\textbf{h}_{k}^{\rm H}\textbf{b}|^2 \beta_{l}^2 + N_0}.
\end{align}

Considering SINRs as in \eqref{eq:SINR_mk_th_user} for all $m$, $ 1 \leq m \leq k-1 $ and SINR for his own data as in \eqref{eq:SINR_k_th_user}, the outage probability at $k$-th user, $\textrm{P}_{k}^{(K)}$ can be given as follows: \begin{align} \label{eq:Outage_k_th_user} 
\textrm{P}_{k}^{(K)} &= 1 - \textrm{P}(\textrm{SINR}_{1({k})} > \epsilon_1^{\rm (T)}, \cdots \ , \textrm{SINR}_{k-1(k)} > \epsilon_{(k-1)}^{\rm (T)},
\nonumber \\ 
&  \hspace{12 em}\textrm{SINR}_{k} > \epsilon_k^{\rm (T)} | K).
\end{align} As captured in \eqref{eq:Outage_k_th_user}, in order to successfully decode his own data, $k$-th user first has to successfully decode and subtract data belongs to $(k-1)$ users with smaller effective channel gains compared to his own channel gain from the observation in \eqref{eq:k_th_user_Observation}. Note here that the outage probability $\textrm{P}_{k}^{(K)}, \, k \in \mathcal{N}_\textrm{U}$ is conditioned on the number of users in the user region, $K$. This is because, $K$ here is a Poisson random variable and observed outage probability is affected from the number of users in the user region.

% Similar to $i$-th user observation in \eqref{eq:k_th_user_Observation}, user $j$'s observation can be given as, \begin{align}\label{eq:j_th_user_Observation}
% y_{j} = \sqrt{P_{\rm Tx}}\textbf{h}_{j}^{\rm H} \textbf{b}(\beta_{i} s_{i} + \beta_{j} s_{j}) + N_0.
% \end{align} 

% \begin{align} \label{eq:SINR_j_th_user}
% \textrm{SINR}_{j} = \frac{P_{\rm Tx} |\textbf{h}_{j}^{\rm H}\textbf{b}|^2 \beta_{j}^2}{N_0}.
% \end{align} 
% The outage probability experienced by user $j$, $\textrm{P}_{j}^{K}$ can then be given as, 
% \begin{align} \label{eq:Outage_j_th_user}
% \textrm{P}_{j}^{K} = 1 - \textrm{P}(\textrm{SINR}_{i({j})} > \epsilon_i^{\rm T}, \ \textrm{SINR}_{j} > \epsilon_j^{\rm T} | K).
% \end{align} 

Considering the outage probabilities as in \eqref{eq:Outage_k_th_user} for all $k \, \in \mathcal{N}_{\textrm{U}}$, the achievable outage sum-rate with the mmWave-NOMA transmission from the UAV-BS can be expressed as: \begin{align} \label{eq:sum_rate_NOMA}
R^{\textrm{NOMA}} &= \textrm{P}(K=1)(1 - \tilde{\textrm{P}}_{1}^{(K)})R_1^{\rm (T)}
\nonumber \\ 
&+ \sum \limits_{n=2}^{\infty} \textrm{P}(K=n) \left( \sum \limits_{k = 1}^{n}  (1- \textrm{P}_{k}^{(K)})R_k^{\rm (T)} \right).
\end{align} where $\tilde{\textrm{P}}_{k}^{(K)}$ is the outage probability of $k$-th $(\in \mathcal{N}_{\textrm{U}})$ user with OMA transmission. NOMA performance is compared with OMA performance and achievable sum-rate with mmWave-OMA transmission from the UAV-BS can be expressed as, \begin{align} \label{eq:sum_rate_OMA}
R^{\textrm{OMA}} &= \textrm{P}(K=1)(1 - \tilde{\textrm{P}}_{1}^{(K)})R_1^{\rm (T)} 
\nonumber \\ 
& + \sum \limits_{n=2}^{ \infty} \textrm{P}(K=n) \left( \sum \limits_{k = 1}^{n}  (1- \tilde{\textrm{P}}_{k}^{(K)})R_k^{\rm (T)} \right).
\end{align} Here, $\tilde{\textrm{P}}_{k}^{(K)} = \textrm{P}\left(\frac{1}{K}\log \left(1 + \frac{P_{\rm Tx}|\textbf{h}_{k}^{\rm H}\textbf{b}|^2}{N_0} \right)< R_k^{\rm (T)}|K \right), \, k \in \mathcal{N}_{\textrm{U}}$. Factor $\frac{1}{K}$ captures the loss of degrees-of-freedom (DoF) gain due to OMA transmission over NOMA.
Note that, as long as required vertical angle satisfies $\varphi_r \leq \varphi_e$, NOMA transmission strategy discussed in this section can be employed. Here after, we consider two user NOMA transmission with $i, j \in \mathcal{N}_{\rm U}$ and $1 \leq i < j \leq K$, since this is the commonly used NOMA transmission approach \cite{Ding17PoorRandBeamforming,Saito13VTC,3GPP16NOMA_LTE}.

% The UAV-BS then picks two users (considering two user NOMA transmission as in \cite{Ding17PoorRandBeamforming,Saito13VTC}),  $i, j \in \mathcal{N}_{\rm U}$ for NOMA transmission, with $i < j$. 
\subsection{NOMA Transmission under Limited Vertical Angle}
\label{Sec:NOMA_Limited_Vert_Angle}
As discussed in Section~\ref{Sec:Behavior_Vert_Angle}, at certain altitudes of some geometries, vertical beamwidth of the UAV-BS beam is not enough to cover the entire user region, i.e., $\varphi_r > \varphi_e$. A beam scanning approach is proposed in Section~\ref{Sec:Beam_Scanning} as a solution to maximize outage sum rates when, $\varphi_r > \varphi_e$.
It is worth remarking that, since our objective is to maximize outage sum rate within the user region by identifying an optimum altitude, when $\varphi_r > \varphi_e$ we modify our NOMA transmission strategy as discussed next without altering the user region.  

Let us define the user set within the radiated region (see Fig.~\ref{fig:footprint}) during beam scanning (when the distance to the boresight intersection point of the beam is $D$) as, $\mathcal{N}_{\rm U}^{D} \subseteq \mathcal{N}_{\rm U}$. Based on the existence of $i, j \ \in \mathcal{N}_{\rm U}$ within the radiated region three possibilities can be identified; case~1: $i, j \in \mathcal{N}_{\rm U}^{D}$, both users are within the radiated region, case~2: $i \in \mathcal{N}_{\rm U}^{D}, j \notin \mathcal{N}_{\rm U}^{D}$ or $j \in \mathcal{N}_{\rm U}^{D}, i \notin \mathcal{N}_{\rm U}^{D}$, only one user is within the radiated region, and case~3: $i, j \ \notin \mathcal{N}_{\rm U}^{D}$, both users are not within the radiated region. 
We alter NOMA transmission strategy for these three cases as follows. For case~1, NOMA transmission discussed in Section~\ref{Sec:NOMA_User_Selection_Data_Transmission} can be directly applied since both users are present within the radiated region. However, when case~2 occurs, the user that falls within the radiated region will be served all the time with full power and outage probability of that user can be given as:
\begin{align} \label{eq:limited_angle_case_2}
\textrm{P}_{k}^{(K)} = \textrm{P}\left(\log \left(1 + \frac{P_{\rm Tx}|\textbf{h}_{k}^{\rm H}\textbf{b}|^2}{N_0} \right)< R_k^{\rm (T)}|K \right)
\end{align} for $ \, k \in \left\lbrace i,j \right\rbrace$ while the other user will be in complete outage. The OMA outage probability is the same as that of NOMA for this case, i.e, $\tilde{\textrm{P}}_{k}^{(K)} = \textrm{P}_{k}^{(K)}, \, k \in \left\lbrace i,j \right\rbrace$. For case~3, no DL transmission will take place and both users will be in full outage. Corresponding outage sum rates in \eqref{eq:sum_rate_NOMA} and \eqref{eq:sum_rate_OMA} for NOMA and OMA, respectively can then be calculated using these outage probabilities for NOMA and OMA.

\section{Numerical Results and Discussion} \label{sec:Numerical_results}

In this section we investigate achievable outage sum rate performance with UAV-BS hovering altitude through extensive computer simulations. In all our investigations, we consider $i=1$ (weakest user) and $j=K$ (strongest user) for NOMA transmission. This type of user selection provides maximum difference in the power domain. Further, we consider two path loss models in our analysis: 1) distance dependent PL as in \cite{Ding17PoorRandBeamforming}, $ \textrm{PL}_k = 1+ d_k^{\gamma}$ where $\gamma$ is the path loss exponent and $d_k$ is the distance between UAV-BS and $k$-th MS, and 2) \emph{close-in} (CI) free space reference distance model for urban micro (UMi) mmWave environment in \cite{CIwhite}, $\textrm{PL}_k = 32.4 + 21\log_{10}(d_k) + 20\log_{10}(f_{\rm c})$. Here, $f_{\rm c}$ captures the operating mmWave frequency. Simulation parameters are summarized in Table~\ref{tab:SimParameters}. 

\begin{table} 
\centering
\caption{Simulation settings.}
\vspace{1 em}
\begin{tabular}{ | c | c | }
\hline
\textbf{Parameter}          & \textbf{Value}  \\
\hline
User distribution          				& Uniformly randomly \\ \hline
Outer Radius, $L_{2}$                   & $100$~m  \\ \hline
Inner Radius, $L_{1}$                   & $25$~m  \\ \hline
Horizontal width, $2 \Delta$            & $0.4^{\circ}$ \\ \hline
Vertical angle, $\varphi_e$             & $28^{\circ}$ \\ \hline
HPPP density, $\lambda$                 & $1$ \\ \hline
%No. of BS ant., $M$               		& 10    \\ \hline
Noise, $N_0$                     	    & $-35$~dBm  \\ \hline
Path loss exp., $\gamma$      		& $2$ \\ \hline
$j^{\textrm{th}}$ user target rate, $R_{j}^{\rm (T)}$    & $6$~BPCU \\\hline
$i^{\textrm{th}}$ user target rate, $R_{i}^{\rm (T)}$    & $0.5$~BPCU \\\hline
$j^{\textrm{th}}$ user power, $\beta_{j}^2$ & $0.25$ \\\hline
$i^{\textrm{th}}$ user power, $\beta_{i}^2$ & $0.75$ \\\hline
Altitude range, $h$             & $10$~m - $150$~m \\ \hline
\end{tabular}
\label{tab:SimParameters}
\end{table}

\begin{figure}[!t]
\begin{center}
\includegraphics[width=0.44\textwidth]{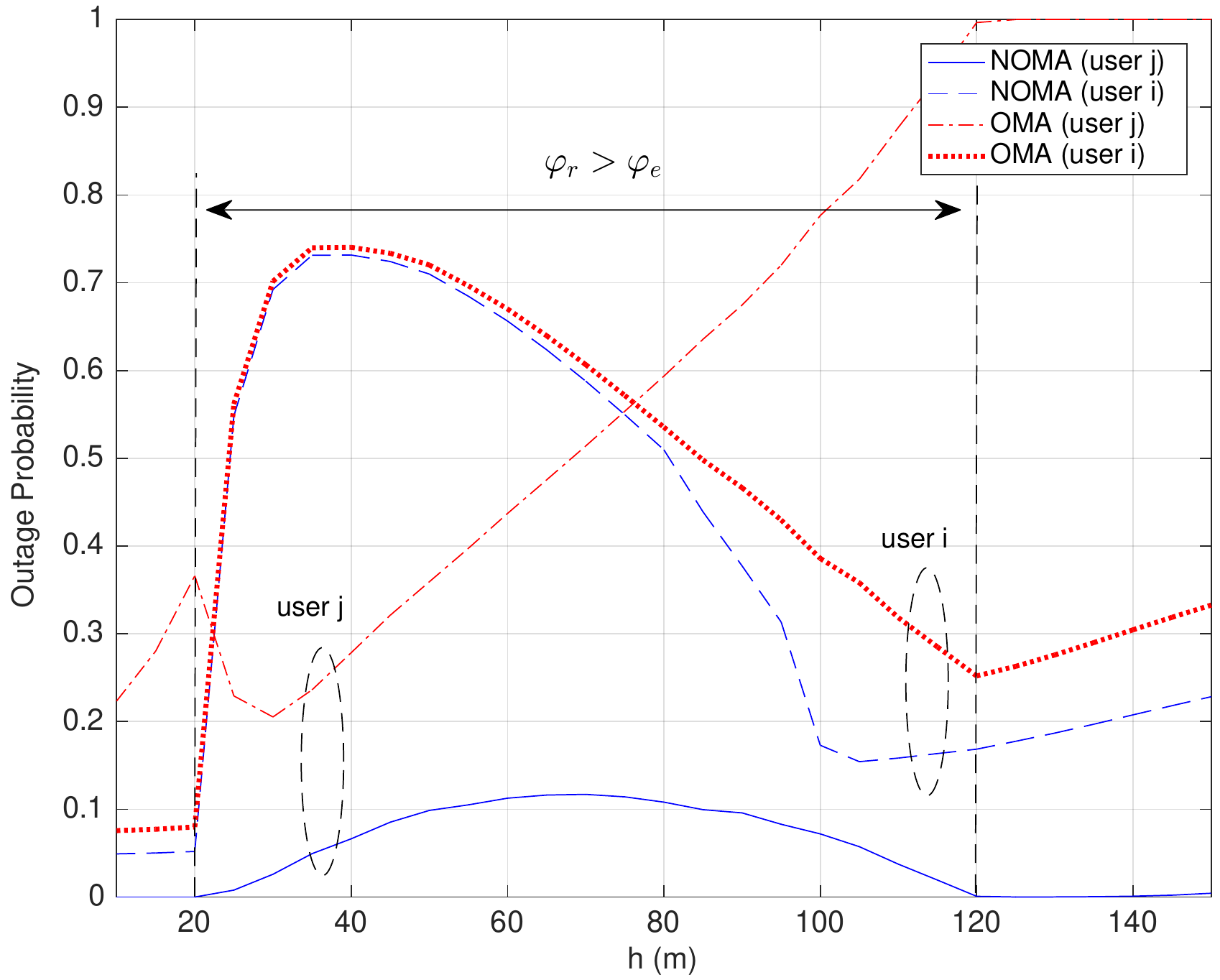}
\end{center}
\caption{Outage Probability variation with UAV-BS hovering altitude for PL model~1 \cite{Ding17PoorRandBeamforming}; $P_{\rm Tx} = 20$~dBm, $M = 10$, and $K = 46$.}
\label{fig:Outage_Probability_P_20_dBm}
\end{figure}

Fig.~\ref{fig:Outage_Probability_P_20_dBm} captures the outage probability variation with UAV-BS hovering altitude. As can be observed, outage probabilities for both $i$-th and $j$-th users are smaller with NOMA transmission compared to that of OMA transmission, at all altitudes of interest. Further, from Fig.~\ref{fig:Outage_Probability_P_20_dBm} it can be observed that, within the altitude range for which $\varphi_r > \varphi _e$, outage probability with NOMA transmission has a convex type of a behavior. This behavior can be explained as follows. As discussed in Section~\ref{Sec:NOMA_Limited_Vert_Angle}, the existence of user $k \in \left\lbrace i,j \right\rbrace$ within the radiated region of the UAV-BS beam is not guaranteed. Hence, there is an associated existence probability for user $k$ at different UAV-BS altitudes when $\varphi_r > \varphi _e$. Intuitively, it can be understood that this existence probability decreases till the peak value of required vertical angle in Fig~\ref{fig:VertAngle} and increases after that. This is because, based on $(\varphi_r - \varphi _e)$ the percentage of the area radiated within the user region changes as shown in Fig.~\ref{fig:VertAngle}. For instance, when the difference $(\varphi_r - \varphi _e)$ is very small, a region closer to the user region can be radiated and hence, there is a higher chance to find user $k \, \in \left\lbrace i,j \right\rbrace $. This existence probability of user $k$ within the radiated region is the dominating factor to observe the convex type of behavior for outage probability at altitudes where $\varphi_r > \varphi _e$. On the other hand, when $\varphi_r \leq \varphi _e$, PL effect reflects in the outage probability performance and if there is sufficient transmit power to overcome PL, outage will be almost zero which is the case for $j$-th user in Fig.~\ref{fig:Outage_Probability_P_20_dBm} with NOMA transmission.

An interesting observation can be made for $j$-th user outage probability with OMA in Fig.~\ref{fig:Outage_Probability_P_20_dBm} when $\varphi_r$ changes from $\varphi_r \leq \varphi _e$ to $\varphi_r > \varphi _e$. This behavior can be explained as follows. When $\varphi_r > \varphi _e$, the beam scanning strategy (maximizing outage sum rate) discussed in Section~\ref{Sec:Beam_Scanning} will tend to radiate a region which includes $j$-th (stronger) user especially at lower altitudes (small PL). Under this situation, when only the $j$-th user is detected (as discussed in Section~\ref{Sec:NOMA_Limited_Vert_Angle}, case~2) the outage probability of the $j$-th user with OMA transmission is given by \eqref{eq:limited_angle_case_2}. This outage probability is better than conventional OMA outage since there is no DoF loss.   

\begin{figure}[!t]
\begin{center}
\includegraphics[width=0.44\textwidth]{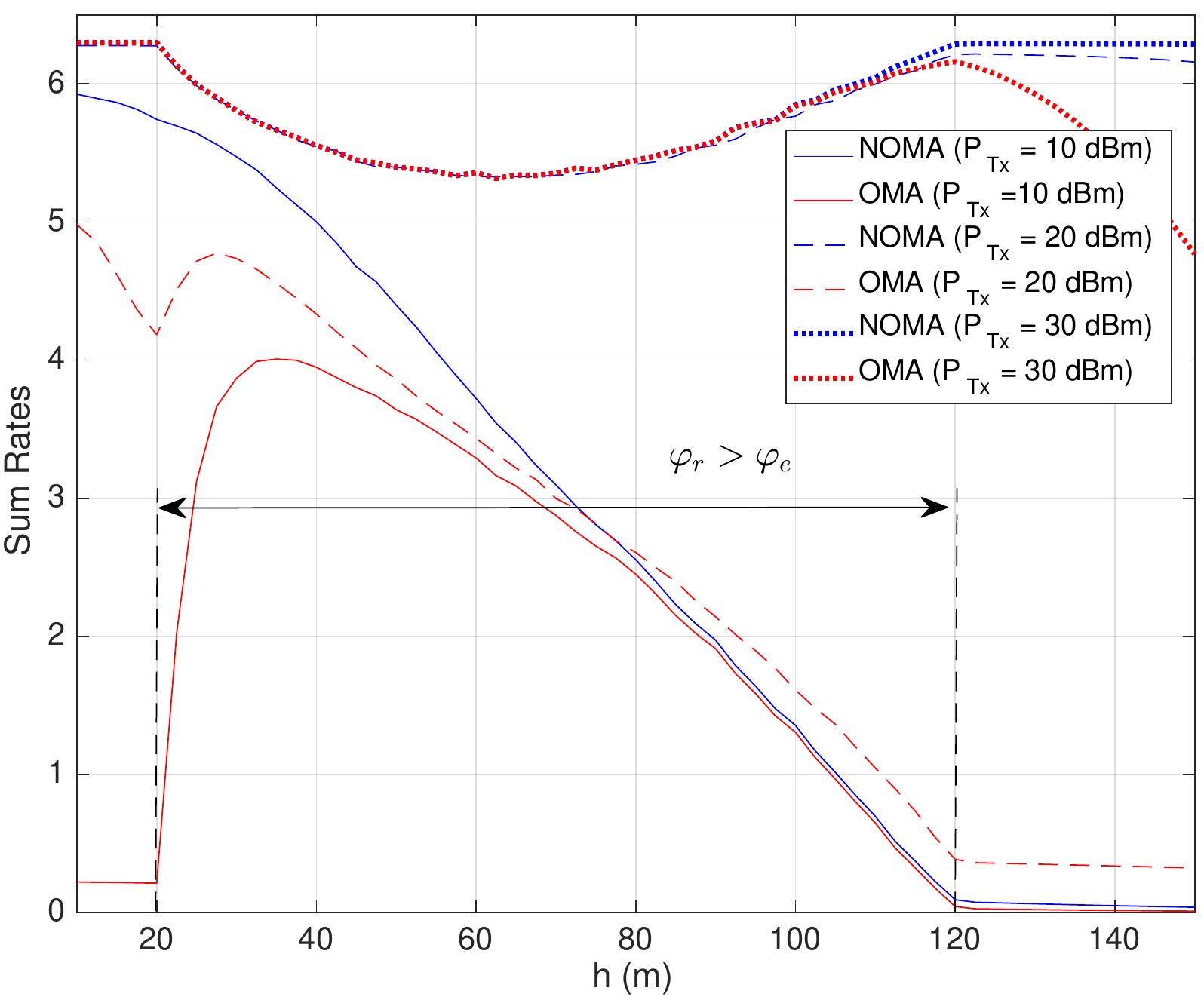}
\end{center}
\caption{Outage sum rate variation with UAV-BS hovering altitude for PL model~1 \cite{Ding17PoorRandBeamforming}; $M = 10$.}
\label{fig:Sum_Rate_Dist_PL}
\end{figure}

Fig.~\ref{fig:Sum_Rate_Dist_PL} depicts outage sum rate variation with UAV-BS altitude for three different transmit power values. It can be observed that outage sum rate values saturate around $6.5$~bits per channel use (BPCU). This is because, as discussed in Section~\ref{Sec:NOMA_Transmission} our focus is to provide targeted rates for users and for this analysis selected target rates are $R_{i}^{\rm (T)} = 0.5$~BPCU and $R_{j}^{\rm (T)} = 6$~BPCU. Hence, sum rate can not go beyond $6.5$~BPCU. Fig.~\ref{fig:Sum_Rate_Dist_PL} shows that NOMA can provide far better sum rate performance than OMA at all altitudes and for all transmit power values. Further the achievable outage sum rates with NOMA for $P_{\rm Tx} = 20$~dBm is same as that with $P_{\rm Tx} = 30$~dBm. This observation suggests that targeted rates can be achieved with $P_{\rm Tx} = 20$~dBm and there is no need to use higher transmit power. This is especially important for UAV-BS communication since UAV-BS has limited power and achieving higher energy efficiency is of paramount importance. 

From Fig.~\ref{fig:Sum_Rate_Dist_PL}, we can investigate achievable outage sum rate behavior for altitudes when $\varphi_r > \varphi _e$. As pointed out in Section~\ref{Sec:Behavior_Vert_Angle}, this altitude range is important since hovering at lower altitudes can raise safety issues and hovering at higher altitudes are restricted due to some regulations. As can be seen from Fig.~\ref{fig:Sum_Rate_Dist_PL}, there is a concave behavior in achievable sum rates when $\varphi_r > \varphi _e$ for higher $P_{\rm Tx}$. This is because outage probability within this altitude range follows a convex behavior as shown in Fig.~\ref{fig:Outage_Probability_P_20_dBm}, mainly due to the existence probabilities of $i$-th and $j$-th users. For smaller $P_{\rm Tx}$, i.e., $P_{\rm Tx} = 10$~dBm, Fig.~\ref{fig:Sum_Rate_Dist_PL} suggests that it is always better to hover at altitudes where vertical angle $\varphi_r > \varphi _e$ in order to maximize outage sum rates rather than flying at higher altitudes where $\varphi_r \leq \varphi _e$, given that flying at lower altitudes are prohibited.  

\begin{figure}[!t]
\begin{center}
\includegraphics[width=0.44\textwidth]{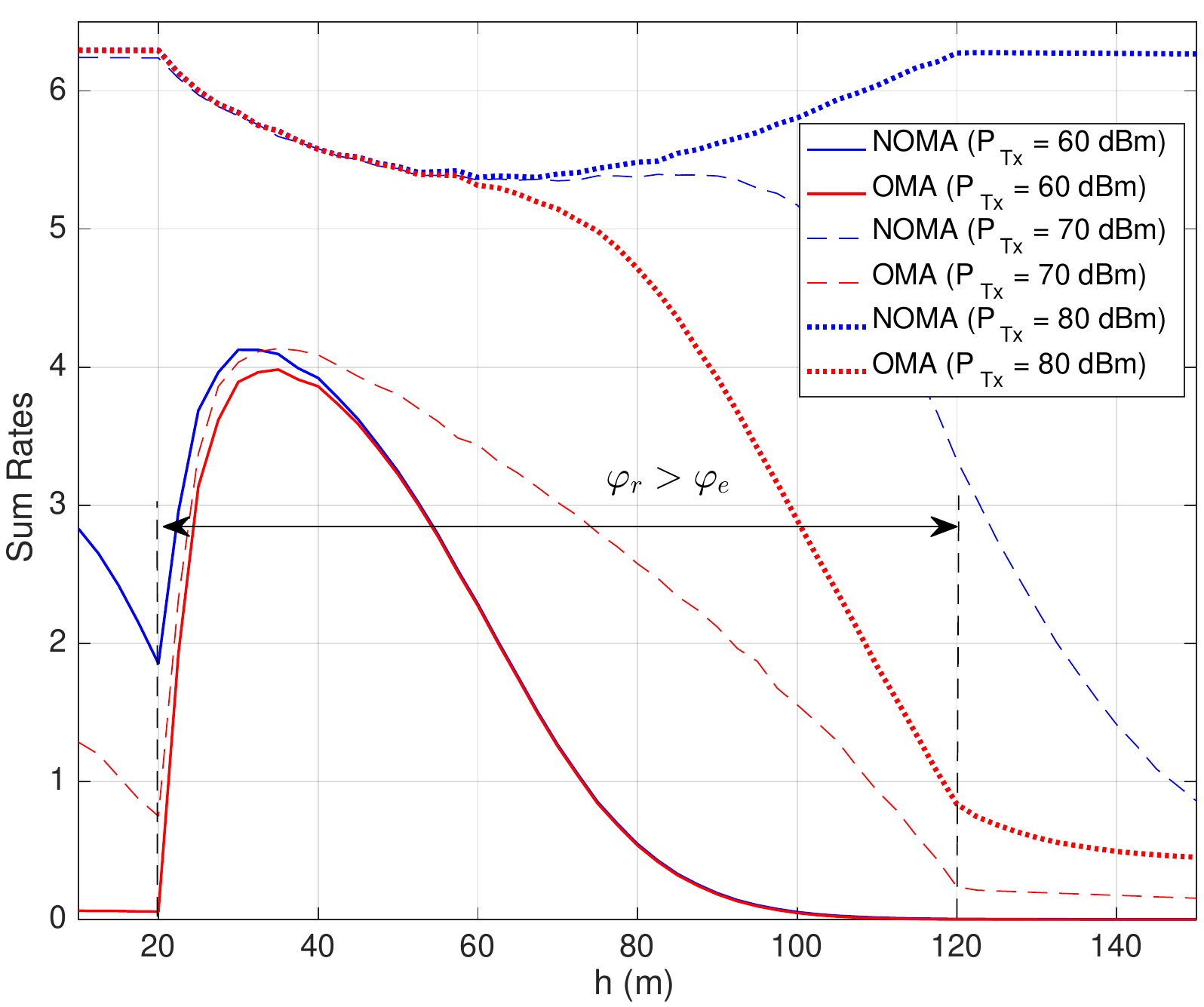}
\end{center}
\caption{Sum rate variation with UAV-BS hovering altitude for PL model~2 \cite{CIwhite}; $f_{\rm c}= 30$~GHz, $M = 100$.}
\label{fig:Sum_Rate_mmWave_PL}
\end{figure}

Finally, Fig.~\ref{fig:Sum_Rate_mmWave_PL} captures outage sum rate performance at different altitudes with mmWave CI path loss model \cite{CIwhite}. For this analysis, we consider $f_{\rm c}= 30$~GHz and $M = 100$. As can be seen from Fig.~\ref{fig:Sum_Rate_mmWave_PL} for the sake of comparison we kept the outage sum rate behavior similar to that in Fig.~\ref{fig:Sum_Rate_Dist_PL}. However, this outage sum rate performance required significantly larger transmit power values ($30$~dB higher than the previous case) and larger beamforming gain. 

\section{Concluding Remarks} \label{sec:conclusion}

In this paper, we investigate how NOMA transmission can be introduced to enhance outage sum rate performance when a UAV-BS is serving during a temporary event such as a concert or sports event. In particular, we focus on maximizing outage sum rate within a \emph{user region} by identifying optimum hovering altitude. The investigation reveals that with NOMA transmission significant improvement in spectral efficiency can be achieved compared to OMA. Further, there is a transmit power value for which target rates for users can always be achieved making it meaningless to increase power any further. 

We study outage sum rate performance within the altitude range at which vertical beamwidth of the UAV-BS beam is not sufficient to cover the entire user region. Hovering at this altitude range is particularly important for UAV-BS since there are various constraints for operating at higher and lower altitudes. The analysis reveals that, when transmit power is small, it is preferable to operate within this altitude range (when flying at lower altitudes are prohibited) even though the entire user region can not be covered.    

\bibliographystyle{IEEEtran}
\bibliography{Doc_Ref}
\end{document}